\documentclass{elsart5p}
\usepackage{graphics}
\usepackage{amsfonts}
\usepackage{graphicx,color,amssymb}
\begin{document}

\begin{frontmatter}

\vspace*{-6.0cm}

\title{R\&D Paths of Pixel Detectors \\ for Vertex Tracking and Radiation Imaging}

\thanks[]{This report has been edited by M. Battaglia and C. Da Vi\'a and includes contributions by
          D.~Bortoletto, R.~Brenner, M.~Campbell, P.~Collins, G.F.~Dalla Betta, P.~Denes, H.~Graafsma, 
          I.M.~Gregor, A.~Kluge, V.~Manzari, C.~Parkes, V.~Re, P.~Riedler, G.~Rizzo, W.~Snoeys and M.~Winter}

\end{frontmatter}

\vspace*{-1.22cm}
\section{Introduction}
\label{sec:1}
\vspace*{-0.5cm}

Semiconductor pixel detectors for vertex tracking have provided the much needed technological breakthrough to make decisive 
advances in the collider exploration of the TeV energy domain and precision heavy flavour physics.
Exceptional progress has been made over the last two decades. Comparable improvements in performance in the next
two decades for upgrades and future colliders are becoming increasingly more difficult. 
Since the design of the LHC detectors, Si pixels have become the standard for vertex tracking in the proximity of 
the interaction region~\cite{Krammer2012}. Applications of pixel sensors beyond vertexing, for the main tracker and high 
granularity calorimetry are also being actively explored, but will not be discussed here.
This report reviews current trends in the R\&D of semiconductor pixellated sensors for vertex tracking and radiation 
imaging. It identifies requirements of future HEP experiments at colliders, needed technological breakthroughs 
and highlights the relation to radiation detection and imaging applications in other fields of science. 
In the following, we consider the LHC luminosity (HL-LHC) and energy (HE-LHC) upgrades and lepton 
colliders at the intensity (superKEKB and SuperB flavour factories) and energy (ILC, CLIC and Muon Collider (MuC)) frontiers. 
There is a broad spectrum of other applications of semiconductor pixel technologies developed towards the specifications 
of these experiments. Their number and diversity is growing, as new facilities put emphasis on specs, such as frame rate, 
radiation tolerance and space-time resolution, which are the primary focus of the HEP-driven R\&D. Examples of the fields where 
pixels developed from results of R\&D for collider experiments include imaging at 
light sources~\cite{Ponchut2002,Henrich2009aa,Dinapoli2011aa} and free electron lasers (FEL)~\cite{Graafsma2009}, including hybrid~\cite{Henrich2011aa} 
and DEPFET~\cite{Porro2010aa} technologies, transmission electron microscopy (TEM), with hybrid~\cite{Faruqi2003,Faruqi2008} and 
monolithic sensors~\cite{Milazzo2005,Battaglia:2010ia}, plasma diagnostics, biological imaging~\cite{Faruqi2009}, auto-radiography, 
with hybrid and monolithic CMOS~\cite{Cappellini2008aa} and DEPFET~\cite{Ulrici2005aa} 
pixels, fluorescence microscopy, with monolithic pixels~\cite{Barbier2011aa}, and medical imaging, with hybrid pixels~\cite{Mikulec2003}, 
beam monitoring and diagnostics for future accelerators, ion-beam analysis~\cite{Pugatch2011} and real-time dose delivery assessment and 
quality assurance in hadron therapy~\cite{Caccia2006aa,Pugatch2012}. There are also examples of returns of these developments 
to the benefit of HEP applications, as discussed in section~\ref{sec:4-3}. In the next chapters we identify some key R\&D areas to enable 
the next generation of collider experiments and then review major trends in semiconductor technology relevant to sensor fabrication, readout 
electronics, services and system-level issues and the availability of test facilities to support the R\&D effort.

To maintain the momentum in the development of pixel systems that will enable an even larger science area to be explored, the key conclusions 
are: 
\begin{itemize}
\item{the concurrent development of hybrid sensors able to sustain high radiation doses and thin monolithic pixels with small cell size is essential;}
\item{microelectronics is key to the success of pixel detectors and investments are necessary to continue the exploration of the benefits from cutting 
edge processes;}
\item{advanced interconnection technologies, including 3D multi-tier, should be strongly supported;} 
\item{system issues are becoming increasingly important and new approaches to integrated detector cooling, such as micro-channels, need to be actively pursued;}
\item{dedicated high rate test facilities are limited and the community should invest in new beam test infrastructure;} 
\item{collaboration with imaging instrumentation development in other field of science, e.g.\ light sources, needs to be encouraged also 
to compensate for the higher investment costs required to access advanced technologies and guarantee broad participation to benefits of the R\&D.}
\end{itemize}

\vspace*{-0.35cm}
\section{Requirements and R\&D Perspectives}
\label{sec:2}
\vspace*{-0.35cm}

The main purpose of Si pixel sensors in collider experiments is the accurate reconstruction of the trajectories of charged particles 
in the proximity of the beam collision point. The standard figure of merit for vertex tracking accuracy is the accuracy on the 
impact parameter, $\sigma_{IP}$, defined as the distance of closest approach of the particle track to the colliding beam position. 
This can be parametrised as: $\sigma_{IP} = a \oplus \frac{b}{p sin^{k} \theta}$, where $p$ is the particle momentum, $\theta$ the 
track polar angle and $k$ = 3/2 for the $R-\Phi$ and 5/2 for the $z$ projection.
The identification of hadronic jets originating from heavy quarks is best achieved by a topological reconstruction of the displaced 
vertex structure and the kinematics associated to their decays. The ability to reconstruct the 
sequence of primary, secondary and tertiary vertices depends on the impact parameter resolution. This is related to the single point 
resolution of the sensors and their geometry. The figure of reference for the resolution is the distribution of the impact parameter 
values for charged particles originating from $b$ hadron decays, which does not depend on the $b$ hadron energy. The average 
value is $\sim$220~$\mu$m, with a 0.15 (0.24) fraction of tracks having impact parameter below 15 (35)~$\mu$m. The asymptotic resolution $a$ 
is typically proportional to the sensor point resolution, the tracking lever arm and, to a lesser extent, the radial position, $R$, of the 
first layer. The multiple scattering term, $b$, scales $\propto \sqrt{d} R / p$ , where $d$ is the material 
thickness. The tracking lever arm is constrained by detector volume, number of electronics channels and cost. The minimum detector 
radius is typically defined by radiation or particle density conditions from particle fluence and beam-induced 
backgrounds and it is accelerator dependant. This was 63~mm at LEP and 25~mm at SLC, is 50-32~mm at the LHC, with plans to further 
reduce it and should be 15-30~mm at future lepton colliders.  

\vspace*{-0.35cm}
\subsection{Radiation Hardness}
\label{sec:2-1}
\vspace*{-0.35cm}

At hadron colliders the location of the detector closest to the collision point depends on the radiation levels and their effects on the 
sensors and readout electronics. The radiation levels for the inner layers of ATLAS, CMS and LHCb upgrades with neutron equivalent fluences 
beyond 10$^{16}$n cm$^{-2}$, or an ionising dose of 1~GRad, significantly degrade traditional sensors of several hundreds $\mu$m thickness 
and require novel approaches. In the case of ALICE, with doses of few 10$^{13}$n cm$^{-2}$ and $\sim$700~kRad, the limitation comes from 
the safe operation of LHC.   
Radiation effects in silicon detectors can be distinguished in two main categories, depending on the type and energy of particles impinging on 
the sensors: ionising and non-ionising effects. Ionising radiation affects the surface oxide layers 
of sensors and electronics with the creation of conductive electron accumulation layers, lateral isolation oxides of MOSFET transistors 
and buried oxides in SOI devices. In sensors, these can be compensated by the deposition, under the oxide, of a p-doping layer 
commonly called p-spray. This effect is dominant in detectors used for X-ray imaging, 
electron microscopy or diagnostic at synchrotron light sources but its effects are present as well in vertex detectors at colliders. 
Non-ionising radiation effects consists in damaging the crystal lattice of the Si bulk and its primary consequence is the generation of defects in the material formed by the combination of vacancies and 
interstitials with dopants and impurities present in its lattice after the sensor processing. When the radiation environment is made of particles 
of different species and energies, it is customary to express the fluence normalised to the equivalent effect of 1~MeV neutrons.

The \underline{radiation levels at hadron colliders} increase with luminosity and decrease with the radial distance from 
the beam.  In the scenarios of energy upgrades, the dose scales with the inelastic cross section within the geometrical acceptance and the 
expected damage increases almost proportional with the beam energy.
Vertex detectors will be exposed to a fluence, mostly due to charged hadrons, ranging from 1-3$\times$10$^{15}$ 
1~MeV neutron equivalent cm$^{-2}$ after an integrated luminosity of 300 fb$^{-1}$ up to $\sim$10$^{16}$ 1~MeV neutron equivalent cm$^{-2}$ 
after $\sim$3000~fb$^{-1}$ at a radius of 32~mm in the general purpose experiments and for the innermost tip of the LHCb VELO upgrade 
detector after 100~fb$^{-1}$.

Concerning the non-ionising radiation effects, many macroscopic bulk parameters of Si detectors change due to irradiation. These include increased leakage 
current and negative space-charge build-up. Space-charge build-up controls the reverse bias voltage needed to generate the $E$ field across the device. 
At high doses, an electric field exists in only some part of the detector thickness 
at reasonable values of the applied bias. Together with charge trapping, this ultimately limits the signal efficiency of the detector. 
The most important parameter for an irradiated detector is the amount of collected charge as a function of the bias voltage, which is affected 
by the increase of the charge trapping due to the radiation induced trapping centres that partially remove charge carriers. The density of traps is 
assumed to increase linearly with fluence. 
The effective drift length, which expresses the length a carrier can drift before being trapped by a defect, is proportional to the carrier's 
mobility and its drift velocity, which reaches a saturation value of $\sim$10$^7$~cm~s$^{-1}$. This length is $\sim$50~$\mu$m for electrons at 
5$\times$10$^{15}$~n~cm$^{-2}$. Operation at high electric fields leads to a consequent reduction of the carrier trapping 
probabilities and consequently to higher signal efficiency.  In addition faster signal is achieved by collecting the electron current which have larger 
mobility. Pixel sensors read out from the n$+$ implant (n-in-n or n-in-p geometries) provide noticeably higher charge collection than standard 
p-in-n detectors after irradiation, as discussed in ~\ref{sec:3-1}. Radiation damage effects on 
the detector operation include: i) depletion voltage, which in traditional planar sensors rises above 600~V and reaches $\sim$1000~V 
for 5$\times$10$^{15}$~n~cm$^{-2}$ in the case of the ATLAS IBL pixel detector layer 32~mm away from the LHC beam. ii) charge trapping which 
reduces the signal. iii) increase in leakage current with 
consequent increase of the noise and the power dissipated by the module. This could trigger a positive feedback effect that forces the module 
into thermal runaway and requires operation well below 0~$^{\circ}$C. iv) reverse annealing, where negative space charge builds up after 
irradiation. This effect can be controlled by oxygenating the silicon bulk and by keeping the sensors below  0~$^{\circ}$C. The cause of 
reverse annealing is presently not completely understood.

At present, the most effective way to improve the radiation hardness of Si sensors is by engineering their geometry to allow 
the highest possible electric field to be applied across their sensitive region. This includes thin and 3D sensors, the latter to be preferred 
in case the front-end electronics (FEE) noise would deteriorate the signal to noise ratio. 3D detectors allow large signal charge to be 
collected for a much lower power dissipation in the sensor, which otherwise would become comparable to that of the read-out circuitry at the 
expected radiation levels. The ATLAS experiment will install 3D detectors in parts of the inner layer during the first LHC long shutdown, and 
obtain valuable operational experience with these devices. Other materials, like diamond, could also be considered for their better noise 
performance, as discussed in~\ref{sec:3-4}. Finally, the charge multiplication phenomenon observed after neutron and hadron irradiation in both 
planar and 3D sensors, interpretable as the onset of a high electric field region where signal charge is multiplied through impact ionisation, 
could be used to the advantage of enhanced detector operation, if it can be controlled. 
An additional challenge specific to the LHCb detector and forward physics experiments is the highly non-uniform distribution of the dose on the detectors, 
inserted in a secondary vacuum perpendicularly to the beam axis, leading to radiation associated annealing effects at one edge of the sensor but not the 
other, while high voltages have to be applied to extract charge from the irradiated side of the sensor. 

The \underline{radiation environment} is radically different \underline{at lepton} \underline{colliders}, where it is the hit density from the beam background 
and not the deposited dose to define the position of the innermost layer. At the ILC, there is an estimated 1.5$\times$10$^{12}$ hits cm$^{-2}$ year$^{-1}$ 
of low momentum background electrons at a radius of 15~mm, corresponding to 150~kRad for a three-year operation period, or a non-ionising radiation 
fluence of $\simeq$2$\times$10$^{11}$ 1~MeV n-equivalent cm$^{-2}$~\cite{Besson:2006aa}. At superKEKB and SuperB, the equivalent fluence is 2$\times$10$^{12}$ 
and 5$\times$10$^{12}$ n cm$^{-2}$ year$^{-1}$, respectively.  These low doses allow the possible implementation of a rich array of technologies which 
cannot be considered in the hadron collider radiation environment. Still, radiation damage needs to be considered, in particular for monolithic pixel 
technologies, which have been 
tested up to doses of several MRad and fluences in excess to 10$^{13}$ n cm$^{-2}$~\cite{Deveaux2007,Koziel2010,Battaglia2010ab,Dorokhov2010,Deveaux2011}. 

These developments extend the applicability of pixel sensors to other fields, such as electron microscopy and imaging at XFELs, where conventional 
imaging devices, such as CCDs, have limitations due to the radiation conditions and read-out speed. In TEM, radiation tolerance requirements depend 
on the mode of operation. For example, operating in diffraction mode, may induce large numbers of electrons to be contained in a limited number of 
Bragg spots, causing large doses to be received over small areas. Given that a very intense bright field image could deposit order of 
10~rad~s$^{-1}$~pixel$^{-1}$, 
a target radiation tolerance of at least a few~MRad enables the use for more than one year. At the European XFEL, the expected dose is quite 
comparable, reaching up to $\sim$3~MRad per year of operations. One important difference compared to HEP applications is the absence, or the 
reduced importance, of bulk damage given the small energy transferred by the soft X-rays and electrons. 

\vspace*{-0.35cm}
\subsection{Space - Time Granularity}
\label{sec:2-2}
\vspace*{-0.35cm}

The two main drivers for the sensor pixel pitch, $P$, are the \underline{single point resolution}, which scales as $P$ and 
the  \underline{local occupancy}, which scales as the pixel area (i.e.\ $\propto P^2$ for square pixels) times the readout time. 
The first pixellated Si device used for vertex tracking in a particle physics experiment was the CCD doublet installed 
next to the target in the ACCMOR experiment at CERN~\cite{ACCMOR} in 1985. It had the pixel density foreseen for the next 
generation of vertex trackers at the STAR HFT at RHIC and at lepton colliders, but on a surface three orders of magnitude 
smaller. This was motivated by both the expected track density next to the target and the required extrapolation resolution 
to study charm decays.
\begin{figure}[h!]
\begin{center}
\includegraphics[width=0.35\textwidth]{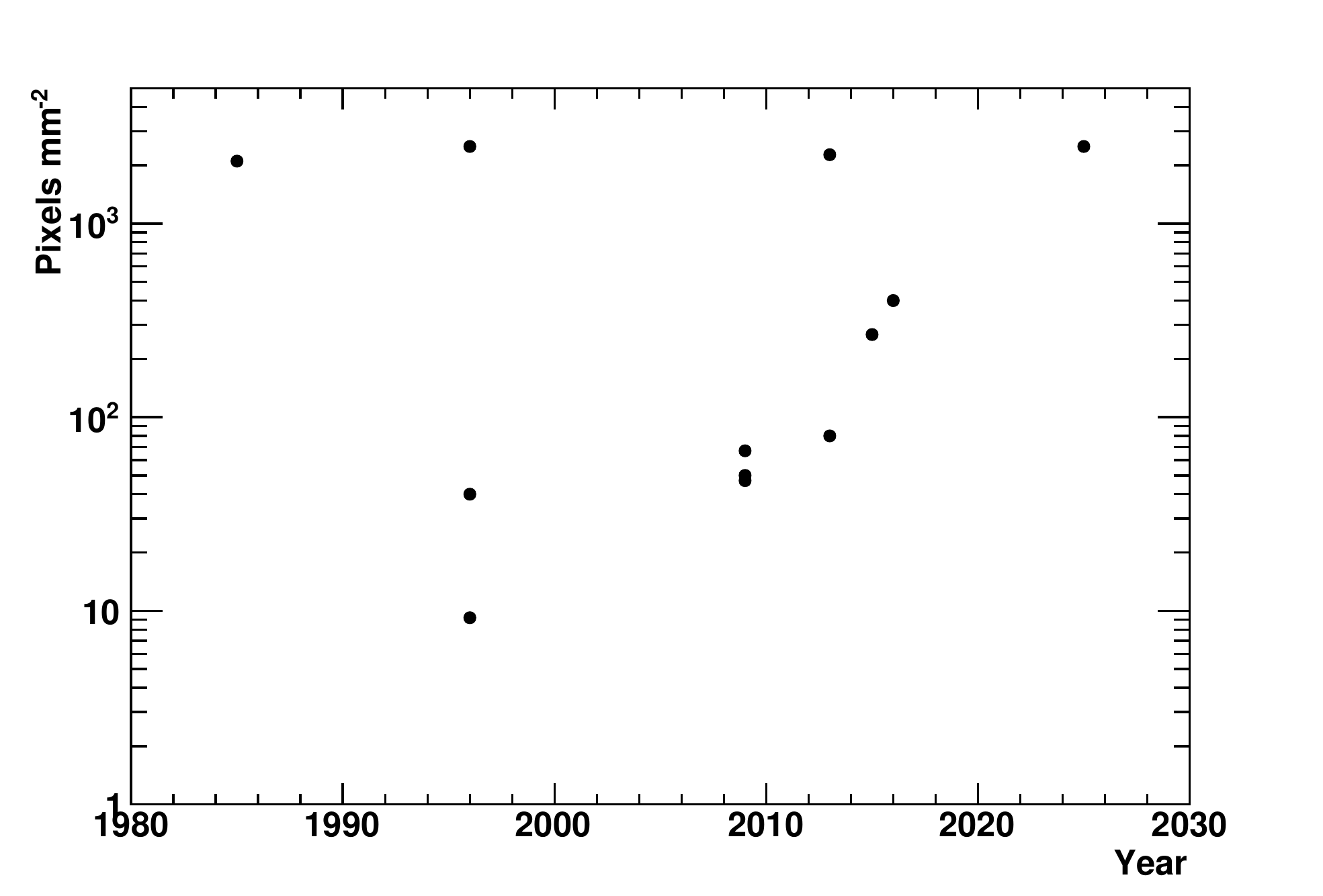}
\end{center}
\vspace*{-0.25cm}
\caption{Evolution of the pixel density vs.\ the year of start of operation of Si pixel vertex trackers. High density detectors 
motivated by point resolution were developed over small surfaces in the early years. The subsequent trend towards increasing 
pixel densities is driven by occupancy and it will match the requirements of high point resolution for future collider projects.}
\label{fig:px}
\end{figure}
At collider experiments, the single point resolution has played so far a major role in determining the pitch of microstrip 
detectors adopted at LEP~\cite{Bingefors1993,Chabaud1996,Mours1996,Acciarri1994,Allport1994} and the pixel size of the CCD 
at SLC~\cite{Abe1997}. With a track density of just 0.004 tracks mm$^{-2}$, the LEP detectors could 
adopt microstrips at a readout pitch of 50~$\mu$m, resulting in an occupancy of $\sim$0.02 and 1D single point resolution 
of $\sim$7~$\mu$m. At \underline{LHC}, where the local track density was estimated to be an order of magnitude larger in $pp$, 
and up to two orders of magnitude in central PbPb, collisions with an average 
$b$-jet energy within a factor of two compared to LEP, the size of the pixel was similar to the LEP pitch in one projection, 
achieving a comparable asymptotic track extrapolation $a \simeq$20~$\mu$m, while the other projection was optimised accounting 
for occupancy and surface available for the front-end electronics and the detector thickness to avoid broken clusters for inclined tracks. 
For the LHCb VELO upgrade~\cite{Collins2011}, the target pixel size of 55$\times$55~$\mu$m$^2$ is still determined by the tracking 
resolution requirements to resolve the oscillations of neutral $B$ mesons, with a decay time resolution of the order of 50~fs. For the 
upgrade planned for 2018 ALICE aims at an $a$ value of $\sim$2.5~$\mu$m, requiring a point resolution of $\sim$4~$\mu$m and an inner 
radius of $\sim$25~mm. The evolution of the pixel density vs.\ year of start of operation is summarised in Figure~\ref{fig:px}. 
Since the typical charge carrier spread is $\sim$7~$\mu$m in a 300~$\mu$m-thick fully depleted pixel and 
$\sim$15~$\mu$m in 15~$\mu$m of an undepleted CMOS sensor, there is limited interest in pushing the pitch much below 10-20~$\mu$m, 
except for imaging applications, where pixels of 5~$\mu$m are used in electron microscopy, or thinner depleted sensors. 
At these sizes, monolithic pixel 
sensors have obtained single point resolutions of 1-2~$\mu$m~\cite{Andricek2011,Battaglia2011aa}, corresponding to an asymptotic 
extrapolation resolution $\simeq 5$~$\mu$m. Track density is expected to become relevant in driving the pixel space granularity.
At LHC displaced decay vertices and high track density in the core of highly boosted jets already lead to merging of hits in the 
pixel layers. The further increase in rate with the LHC upgrades and the short bunch spacing, or large continuous background flux, 
at high energy lepton colliders, such as CLIC and a MuC respectively, push the design of the sensors to be located 
closest to the interaction region towards smaller cells, driven by both occupancy and two-track separation, which again scales 
$\propto P$. At a \underline{linear collider}, both background and physics track density are important. The innermost radius of the 
vertex tracker is determined by the density of incoherent pairs produced and deflected in the e.m.\ interaction of the two 
bunches. Two-track separation is challenged by particle density in the core of boosted hadronic jets where the minimum distance 
between two contiguous hits on the first layer of the vertex tracker has a most probable value of $\sim$400~$\mu$m in 
$e^+e^- \rightarrow H^0Z^0 \to b \bar b q \bar q$ events at $\sqrt{s}$ = 500~GeV  and just $\sim$240~$\mu$m in 
$e^+e^- \rightarrow H^0A^0 \to b \bar b b \bar b$ events at $\sqrt{s}$ = 3~TeV. Typical two-track separation is 2-3$\times$P, 
depending on the cluster size. In the case of \underline{flavour factories} the advantages of pixels over striplets are found 
in the more stable performance in high background scenarios thanks to their lower occupancy and are therefore the technology of 
choice for BELLE-II and SuperB at full luminosity. A study performed for SuperB compares the physics sensitivity in time-dependent
analyses normalised to the number of reconstructed physics events for striplets and pixels of equal thickness and resolution.
The relative degradation of the sensitivity for pixels is a factor five lower compared to that for striplets, assuming a 
background rate up to five times the simulation expectation to include a safety factor.
Starting with the current LHC detectors, fast time-stamping, required by occupancy, has become part of the 
overall optimisation and we have to consider the space-time granularity as the appropriate parameter space for 
any future vertex detectors. New approaches, such as ultra-fast timing exploiting charge multiplication in thin sensors, are being 
investigated. Fast \underline{time stamping} is of special importance in occupancy reduction at accelerators with long trains of 
closely spaced bunches, such as a linear collider. This has obvious benefits beyond HEP for applications such as fourth generation 
\underline{light sources and FELs}. In these applications there is comparable interest in reducing the pixel pitch from the current 100-500~$\mu$m 
down to $\sim$50~$\mu$m or even less. The burst mode time structure of the European XFEL imposes to store images in the detector front-end 
during the full bunch train for subsequent readout during the inter-train period. The area required for memory puts severe limits on the minimum 
pixel size that can be achieved. This limitation will be eased by the reduction in feature size of the CMOS processes.

\vspace*{-0.35cm}
\subsection{Sensor Thickness}
\label{sec:2-3}
\vspace*{-0.35cm}

\begin{figure}[h!]
\begin{center}
\includegraphics[width=0.35\textwidth]{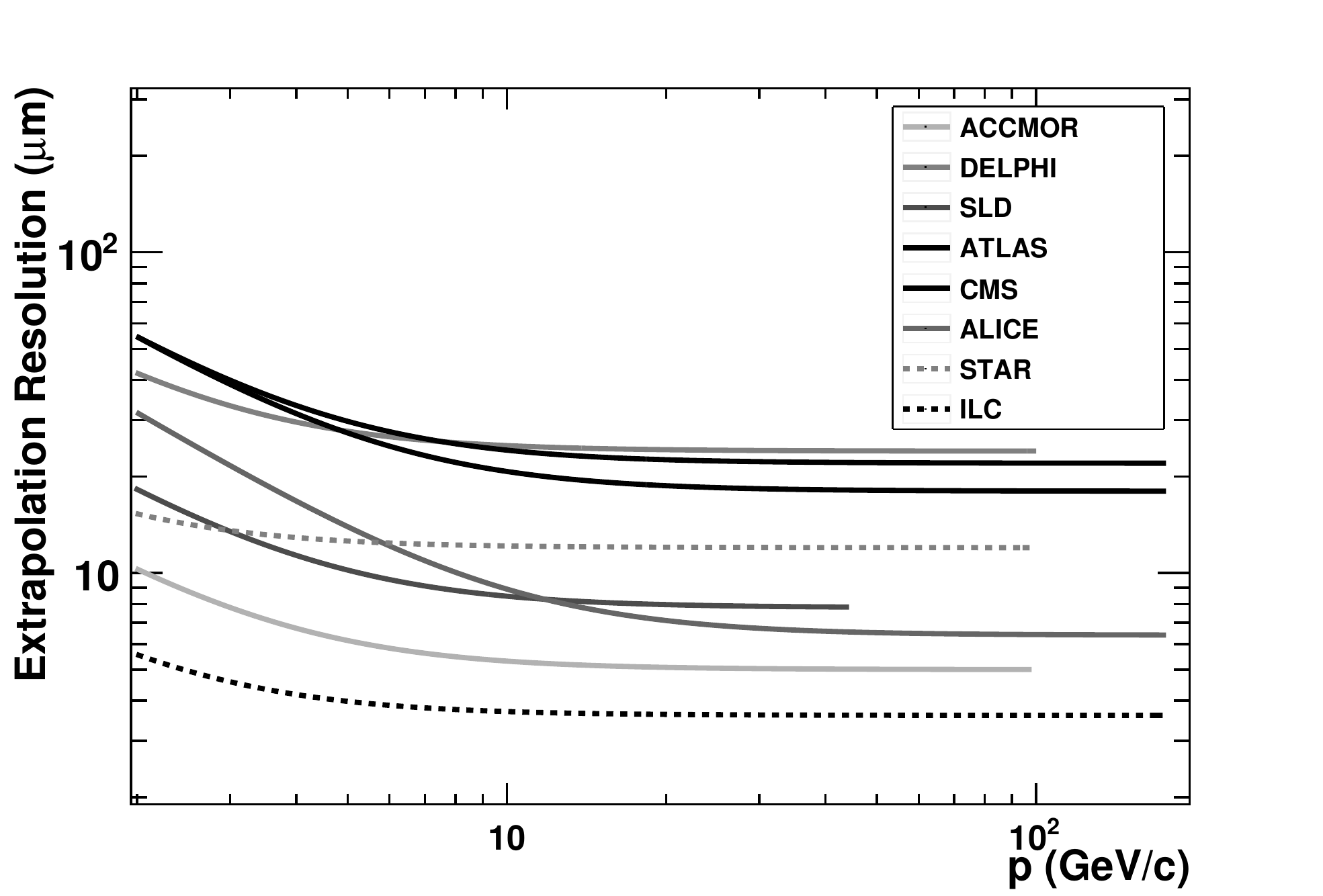}
\end{center}
\vspace*{-0.30cm}
\caption{Extrapolation resolution vs.\ track momentum for past, present and future Si vertex detectors.}
\label{fig:ip}
\end{figure}

In collider experiments a reduction of the detector thickness is highly desirable because it minimises the amount of 
multiple scattering experienced by charged particles, in particular in the first layer(s) of the vertex tracker, as well 
as the number of radiation lengths of material placed in front of the calorimeters. This has implications 
on the precision of extrapolating (mostly soft) particles back to their production vertex and on the amount of energy loss
by radiation before the calorimetric measurements. The track extrapolation resolution for several Si vertex detectors is 
given in Figure~\ref{fig:ip}. 
It is interesting to observe that even at multi-TeV collision 
energies there is a sizable fraction of final state hadrons which do not exceed a few GeV in 
momentum. As an example, the fraction of charged particles originating from a Higgs boson decay to a pair of $b$ quarks and 
having a momentum below 2.5~GeV/$c$ ranges between 0.24 and 0.19 for $pp$ collisions from $\sqrt{s}$ = 8~TeV to 30~TeV and 
between 0.26 and 0.19 for $e^+e^-$ collisions from $\sqrt{s}$ = 0.5~TeV to 3~TeV. At \underline{flavour factories} and in 
\underline{heavy ion collisions}, soft particles from heavy hadron decays are dominant.  Thinner sensors are also less susceptible 
to $\delta$-ray emission, resulting in a displacement of the collected charge w.r.t.\ the particle point of impact. Sensor 
thickness accounts for only part of the total material budget of vertex tracking systems, with support structures, cooling and 
other services often being the dominant contribution. In the sensitive part of the ATLAS pixel detector the sensors with their 
readout electronics account for 17\% of the total material, the largest contribution being due to the mechanical structure and 
cooling components at 58\%~\cite{Gonella2011aa}. The present ALICE pixel detector is the lightest vertex detector among the LHC 
experiment with a total material budget of 1.14~\% $X_0$, where the contribution of the sensor and FEE chip amounts to 33\%. 
Still, the availability of thin sensors is important to precision vertex tracking. 
In particular, this is the case in experiments where high flavour tagging efficiency, such as high-energy lepton colliders, or 
reconstruction of low momentum short-lived hadrons, such as $b$-factories and heavy ion experiments, is of paramount importance 
to their physics program. Thin sensors are also valuable to imaging applications, for example in TEM where the reduction 
in multiple scattering improves the point spread function (PSF). 

The practical thickness is tightly linked to the amount of signal charge collected for a minimum ionising 
particle. This has to be compared to the intrinsic single pixel noise in terms of the achievable signal-to-noise ratio (S/N). 
The S/N determines the detection efficiency, taking into account charge sharing on neighbouring pixels and discriminator threshold 
setting, and also the single point resolution, if analog read-out with charge interpolation is used.
The efficiency starts to degrade for S/N values below 20 per pixel. A minimum ionising particle generates in Si a most probable value 
of $\sim$70~e$^-$ $\mu$m$^{-1}$. With a typical noise of less than 200~e$^-$ ENC and taking into account the minimal threshold of the 
readout chip which allows stable operation, a 250-300~$\mu$m thick \underline{hybrid pixel} sensor, 
as those employed in the ATLAS and CMS vertex trackers, the resulting S/N ensures a comfortable margin for detection efficiency, 
but imposes a multiple scattering burden of almost 2\% $X_0$, when also the readout chip is accounted for. This is becoming more important 
for the next generation of vertex trackers. The reduction in material, obtained by shifting inactive components out of the sensitive region 
and reducing the contribution of cooling and flex cables, in the upgrades brings the attention back to the active components. In CMS, 
the sensor+readout-chip will contribute 56\% of the material of a single ladder, up from 42\% for the current detector. 
The readout chip contributes significantly to the material budget and a main limitation to its thinning has been the bending at the 
high temperatures used for re-flow during the flip chip process. In the case of the current FE-I3 ATLAS pixel chip, with an area of 
0.8~cm$^2$, a thickness of 190~$\mu$m is adequate, but the 4~cm$^2$ FE-I4 chip had to be supported by a handle chip substrate to safely 
bond it at a thickness of 150~$\mu$m. New flip chip techniques have been developed for the next generation of hybrid pixel modules for 
LHC upgrades to reduce the chip contribution to the material budget~\cite{Gonella2011aa}.

Much thinner active layers can be afforded by \underline{monolithic} CMOS and DEPFET technologies, introduced in the R\&D for the 
linear collider, where the pixel cell measures ${\cal{O}} \mathrm{(10 \times 10)}$~$\mu$m$^2$ with a typical capacitance of 
${\cal{O}} \mathrm{(1)}$~fF and 10-40~e$^-$ ENC of noise. An active thickness of 15-20~$\mu$m is sufficient for achieving a detection 
efficiency $\ge$99~\% even at operating temperatures in excess to 40$^\circ$~\cite{Winter2010aa}. 
Back-thinning technologies available as commercial service, such as grinding and chemical etching, have been successfully applied 
to make 40-50~$\mu$m-thick CMOS sensors for the STAR HFT at RHIC~\cite{Greiner:2011aa}, the R\&D for the linear collider~\cite{Battaglia2007} 
and Super-B~\cite{Rizzo:2011zz} vertex detector. The ALICE upgrade baseline design is also based on monolithic pixels for an overall material 
budget 0.3-0.4~\% $X_0$ / layer with the sensor contributing $\sim$20\%.  
DEPFET sensors can be thinned to 50~$\mu$m in the sensitive region, using a wafer bonding technology, retaining a frame which ensures the stiffness 
of the mechanical module~\cite{Andricek2004}. Thin DEPFETs are under development for 
BELLE-II~\cite{Fischer:2007zzm,Marinas2012aa}. Sensors of this thickness are compatible with the needs of $b$-factories and for achieving a material 
budget of $\sim$0.1~\%~$X_0$/layer, needed to meet the value of $b \sim$10~$\mu$m for the multiple scattering term, identified as a requirement 
at a linear collider. 
For sensors of this thickness, the ladder support is of major importance for counteracting the chip warping and insuring the module 
stability and planarity. In the design of thin ladders for applications at $b$-factories, STAR and the linear collider with thin 
sensors, as well as those for an LHC upgrade, the Si ceases to be the largest single factor in the overall layer 
material budget which becomes the cable routing signals, power and clocks. In order to progress toward thinner modules, sensor 
stitching, with clocks and signals routed on metal lines in the chip, appears a promising path. 

Progress with thin sensors benefits \underline{other applications} involving soft radiation, where multiple scattering and 
$\delta$-ray production is an issue. In TEM, replacing phosphor-coupled CCDs with direct detection on CMOS sensor is the latest trend 
in imaging~\cite{Milazzo2005,Deptuch2007ab,Battaglia:2010ia}, to decrease the PSF 
and enhance the detection quantum efficiency and imaging contrast ratio. Multiple scattering in the sensor is a major 
limitation, since the typical electron energy is in the range 60 - 300~keV. CMOS sensors, with a $\simeq$15~$\mu$m-thick sensitive volume, 
thinned to 50~$\mu$m improve the contrast ratio by a factor 1.5 and the PSF by 1/3 compared to 300~$\mu$m-thick sensors of the same 
design~\cite{Battaglia:2010ia}. Thin monolithic sensors are well suited for the detection of soft X-rays and charged particles 
in applications from XFEL to solid-state dosimetry by thinning the sensor to the boundary of the sensitive layer (depleted or epitaxial) and 
using it in back-illumination configuration~\cite{Deptuch2007,Boll2011,Battaglia2012az}. Thinning is important to open transparent windows 
to soft radiation but the thickness of the readout chip is not limited in these applications as for vertex tracking. Therefore the possibilities 
offered by the 3D ASIC developments, discussed in section~\ref{sec:3-3}, can be fully exploited. 

\vspace*{-0.35cm}
\subsection{Energy Resolution}
\label{sec:2-4}
\vspace*{-0.35cm}

Typical HEP applications do not place emphasis on the resolution of the energy deposited in the pixels. 
However, there are examples of the use of dE/dx measured in the Si trackers for improving momentum resolution~\cite{Paganis2001} and performing 
particle identification~\cite{WW} at low momenta, also applicable to proton tomography~\cite{pCT}. Pixellated semiconductor sensors may offer 
energy resolution comparable to that of commercial spectroscopy detectors with the added benefits of spatial resolution. 
Fully depleted CCD on high resistivity substrate~\cite{Holland2003aa}, developed mostly for astronomical applications, have 
an energy resolution of $\sim$150~eV FWHM at 5.9~keV and are successfully applied in imaging and XRF analyses at light 
sources and for plasma diagnostics~\cite{Thorn2010}. Similarly, pnCCDs have demonstrated excellent energy resolution and are used 
at LCLS~\cite{Struder2010}. The very low input capacitance and in-pixel charge to current conversion 
of monolithic pixel sensors are advantageous in terms of pixel noise and makes them attractive for applications requiring 
good energy resolution at relatively high frame rate. 
DEPFET sensors, derived from R\&D for the ILC vertex tracker~\cite{Velthuis2007aa}, have demonstrated single-pixel energy 
resolution of 130~eV at 5.9~keV, which is Fano limited. They are being developed for use in satellite missions for X-ray 
astronomy~\cite{Stefanescu2010aa} and planetary observation~\cite{Treis2010aa} and X-ray light sources~\cite{Porro2010aa}. 
SiC pixels have obtained an energy resolution of 196~eV at 5.9~keV~\cite{Caccia2011aa} and CMOS LePix sensors gave 
143~eV~\cite{Mattiazzo2012}.
Pixels in SOI technology are also being developed towards applications in X-ray detection. Their noise performance is 
not yet a match to the examples above, due to the larger pixel capacitance and leakage. However, there are promising 
R\&D avenues to make these sensors applicable to X-ray astronomy~\cite{Ryu2011aa}. In addition, 
back-plane post-processing to create a thin entrance window, offering enhanced spectral sensitivity to soft X-rays has 
been demonstrated~\cite{Battaglia2012az}. 

\vspace*{-0.35cm}
\section{Semiconductor Technology Trends}
\label{sec:3}
\vspace*{-0.35cm}

Recent developments in the field of microelectronics are leading to a performance leap for pixel sensors, thanks to 
an aggressive scaling of the feature size of CMOS technologies. 
Commercial silicon foundries have been able to offer a density advantage of a factor of two 
at every 18-month generation, and are now capable of offering high volume technologies with features well below 
100~nm, with the main goal of increasing performance and density of large digital systems. The industrial road-map 
foresees the introduction of processes below 15nm by 2015, accompanied by the further development of many special 
technologies, such as vertical integration, integration of special features such as bipolar, germanium-doped or 
power devices, SOI devices and many others. Microelectronic design 
based on CMOS scaling has to tackle bottlenecks, such as the length of interconnections. Many of these have been already been 
overcome through engineering creativity and, of course, investments. In future, for thin oxides gate leakage and voltage swing 
become a concern, particularly for monolithic CMOS active pixels. 3D integration of two or more vertically interconnected layers 
of CMOS circuits promises to help by shortening connection paths and by increasing functional density~\cite{Yarema2010,Lipton2011}. 
3D techniques can be exploited in the fabrication of multilayer electronic devices, where each layer is based on a different 
technology and optimised according to its function (sensing, analog and digital signal processing, storage and data transmission).

The CMS phase-1 upgrade~\cite{Favaro2011} is still in 250~nm technology but 65~nm CMOS is foreseen for the phase~2 and the development 
of readout chips for the LHCb~\cite{Collins2011} and ATLAS upgrade~\cite{Barbero2011} are in 130~nm technology. The design of complex 
embedded trigger logic for more intelligent vertex trackers are in 65~nm and with obvious advantage of high-speed logic such as that 
needed in fast data links~\cite{Gaioni2011b}. Validating such technologies requires assessing their reliability in high radiation environments 
and great progress has recently been made in proving the robustness of commercial 65~nm CMOS~\cite{Gaioni2011}. Nonetheless, still more work 
will be necessary and should be funded to assess the potential of 45nm and below. The time and costs of developing and perfecting new 
technologies are felt to be a limiting factor. The boom and bust cycle, which dominates particle physics in terms of long 
development times alternating with experiment construction, leads to great difficulties in developing performant ASIC necessary 
for future experiments. To this end experiments and R\&D collaborations support the diversification of applications 
between particle physics, medical applications, instrumentation of beam lines, surveying detectors and so on, in order to make the 
incremental steps necessary to master new semiconductor technologies.  

\vspace*{-0.35cm}
\subsection{Sensors}
\label{sec:3-1}
\vspace*{-0.35cm}

Fabrication technologies for \underline {planar pixel sensors} have been well established for several decades. P-on-n pixels were first introduced 
and are the simplest to fabricate, since they require a single side technology without need for surface isolation layers between p-type pixels. 
As an example, they are currently used in the ALICE experiment at the LHC~\cite{Riedler07}. P-on-n pixels have also been fabricated on epitaxial 
silicon, an interesting solution in case thin active layers are required~\cite{Calvo10}, with promising features of higher radiation 
hardness, as demonstrated by studies on pad detectors~\cite{Lindstroem06}.   
The two main options currently available for planar pixels are those featuring n-side signal read-out, i.e., n-on-n and n-on-p,
to exploit the higher mobility of electrons with respect to holes, which results in faster signals and higher radiation tolerance~\cite{Affolder10}. 
N-on-n pixels were developed in the 90's and were finally the sensors choice for the ATLAS and CMS pixel detectors~\cite{Huegging01,Allkofer08}. 
The entire pixel side (front side) is kept at ground, so there is no risk of micro discharges between the sensor and the read-out chip after bump 
bonding. Isolation between n-type pixels is achieved by surface p-type implantation (p-stop, p-spray or moderated p-spray)~\cite{Richter96,Piemonte06}. 
The back side must  also be patterned to have a sufficiently wide gap (often hosting floating guard rings) between the main junction where the 
reverse bias voltage is applied and the cut edge of the sensor, that must be at ground. Thus, a double side fabrication process is required, 
with impact on costs. N-on-n pixels with slim edge (200~$\mu$m) made on 200~$\mu$m thick substrates have been recently adopted for the production 
of the planar pixels for the ATLAS IBL~\cite{Klingenberg12}. 

An interesting and less expensive alternative are n-on-p sensors, that need a simpler, single-side technology (only the pixel side is patterned). 
This option has been considered since p-type detector grade silicon has become routinely available and n-on-p 
pixel sensors are currently developed at different facilities in view of the upgrades at the LHC~\cite{Affolder11}. Comparable performance have 
indeed been observed for n-on-n and n-on-p sensors in terms of charge collection efficiency after large radiation fluences~\cite{Affolder10}. 
A disadvantage of n-on-p sensors is that the edge on the pixel side is at same potential as the bias and can be very high for irradiated devices.
To avoid risk of micro discharges between sensor and read-out chip, special passivation treatments of the sensors are required (e.g., 
a benzocyclobutene layer~\cite{Macchiolo08}). Another disadvantage with the "n-on-p" approach is the lower design flexibility for reducing the 
dead area at the edge, because the guard rings are on the same side as the pixels. The minimum edge so far achieved with "n-on-p" pixel sensors is 
450~$\mu$m~\cite{Gallrapp12}. 

\underline{3D pixel sensors} \cite{Parker97} have been the object of extensive R\&D in the past years and will play a major role in future experiments 
at large luminosity colliders. The successful fabrication of more than 300 3D sensors compatible with the FE-I4 read-out chip for the 
ATLAS IBL has also demonstrated that 3D sensor technology is suitable for medium volume productions \cite{DaVia12}. 

Two different 3D processing options exist, single-side and double-side. The first was originally fabricated at SNF (Stanford U., CA, USA)~\cite{Kenney99} 
and is now also available at SINTEF\footnote{SINTEF, Oslo, Norway.}~\cite{Hansen10}: column etching is performed by DRIE (Deep Reactive Ion Etching) 
all through the substrate from the front side of the sensor wafer for both types of electrodes, at the same time implementing active ohmic trenches 
at the edge, whereas the bottom side is oxide bonded to a support wafer. This requires extra steps to initially attach and finally remove the support wafer. 
The second option is a double-side process, independently developed by CNM (Barcelona, Spain) and FBK (Trento, Italy)  in slightly different versions: 
junction columns are etched from the front side, ohmic columns from the back side, without the presence of a support wafer. In CNM sensors, columns 
do not pass through but stop at a short distance from the surface of the opposite side~\cite{Pellegrini08}. This was also the case for the first 
prototypes of FBK sensors~\cite{Zoboli08}, but later the technology was modified to allow for passing through columns.
Also for 3D sensors, both single-side and double-side, electron signal collection is preferred, and n-type pixels are isolated by surface p-type 
implantation (p-spray or p-stop). 
Double-side 3D sensor fabrication is simpler and faster because it avoids the support wafer and related process steps, with the additional advantage 
of having the back-side accessible to apply the bias voltage. On the other hand, in the absence of the support wafer, active edges cannot be obtained. 
As an alternative, slim edge solutions have been implemented, based on 3D guard-rings~\cite{Pellegrini12} or ohmic fences~\cite{Povoli12}, yielding 
a dead area at the edge in the order of 200$\mu$m or lower for optimised designs.

Future directions for pixel sensors involve lower material budget (lower thickness), very slim edges and higher radiation tolerance. On the 
fabrication side, substrate thicknesses in the order of 150~$\mu$m are still suitable for production with standard equipment. Lower thicknesses can 
be obtained either by local thinning of the substrate~\cite{Ronchin04} or by using a support wafer which is later removed~\cite{Andricek04}. 
As an alternative, in epitaxial wafers the heavily doped substrate can act as an intrinsic support wafer and can finally be almost completely 
removed~\cite{Calvo10}. In all these cases, the possibility to pattern the back-side is limited if not absent, in case of epitaxial.
Thin planar sensors involve dealing with higher capacitance (noise) and lower signals. An interesting solution to improve the S/N ratio would be to 
exploit charge multiplication processes, as those observed in heavily irradiated strip sensors with n-side read-out biased at very large voltages 
($\simeq$1000~V)~\cite{Casse10}. Ideally, charge multiplication should be controlled by design and take place at lower voltages. Special designs 
involving additional implants or other structures enhancing the electric field are presently under consideration~\cite{Casse12}.

As for the \underline{very slim edge} objective, there are a few possible approaches. Scribe-Cleave-Passivated (SCP) technique is being investigated, involving 
wafer scribing and cleaving to produce fewer defects than standard dicing saw, and edge surface passivation, e.g., by means of an alumina layer, 
that is suitable for p-type substrates due to its negative fixed charge~\cite{Christophersen12}. This method could push the dead area at the edge down 
to $\le$100~$\mu$m and it can be applied as a post processing step regardless of the specific technology. It has already been proved with good results 
on few samples using a manual equipment but still needs to be extensively tested after irradiation and fully automatised for volume production. 
The process of stealth dicing based on a laser generating a stress layer inside the wafer can provide even narrower edges\footnote{Hamamatsu Photonics, Japan}.

Planar pixel sensors with active edge, derived from the original 3D sensor technology~\cite{Kenney01}, might offer the best performance~\cite{Kenney06}, 
with full signal sensitivity up to a few microns from the physical edge, but the radiation tolerance limits of these devices have not yet been assessed. 
On the technological side, the main issue is the support wafer removal step that is very challenging in the presence of etched trenches, and still 
needs to be properly engineered. In addition, to ease assembly, the insulating layers on the back side should also be locally etched and a metal 
should be deposited in order to apply the bias voltage. In this respect, using epitaxial wafers might offer an advantage, because a good ohmic 
contact could be made on the back side directly depositing a metal over the remaining substrate layer after its etching.

Similar considerations apply to 3D sensors: using the support wafer, the single side option comes with the active edge and can easily be scaled to 
lower active thickness, that would actually ease the electrode etching and filling steps, thus allowing for a significant process simplification. 
Narrower electrodes could also be obtained, with better signal efficiency and lower capacitance. Again, support wafer removal is the main concern, 
and could be even more difficult for thin active layers. Conversely, for double-side 3D sensors, reducing the thickness below 200$\mu$m will be very 
complicated, particularly in terms of wafer breakage risk during processing. As for the slim edge, novel termination designs have been proposed that 
could allow for a $\simeq$50~$\mu$m dead area, but also the SCP approach could provide a good solution.

Charge multiplication has also been observed in irradiated and non-irradiated 3D sensors at voltages of about 200~V~\cite{Koehler11}, and, if 
properly controlled, would be appealing for thin 3D sensors, aimed at compensating signal reduction due to the thinner substrates and to charge 
trapping after heavy irradiation. The feasibility of charge multiplication after irradiation depends on the high electric fields close to the 
junction electrodes because of the large density of radiation-induced traps, further enhanced by tip effects for non-passing-through columns, 
before irradiation. It is unlikely that charge multiplication might occur unless very high voltages are applied. In fact, high electric fields can 
also be obtained by adopting new designs with closer electrode spacing, that would become feasible in case of thin substrates with narrow electrodes. 
TCAD simulations show promising results with charge multiplication gain of up to about seven at reasonably low voltages (100-120~V).

Two dimensional pixellated detectors have long played crucial roles in \underline{photon science}. Phosphor screens, fiber-optically coupled to CCDs have been
work horse X-ray detectors. Increasingly, these are becoming replaced with direct detection, semiconductor detectors. Hybrid pixels are becoming ubiquitous for 
hard X-ray applications, such as macromolecular crystallography. Originally used as photon counters, whereby a incident X-ray would fire a discriminator
and increment an in-pixel counter, integrating pixels are also finding application, particularly with free electron lasers (FELs). 
The PILATUS~\cite{Henrich2009aa} detector system, and its successor EIGER~\cite{Dinapoli2011aa}, for imaging in synchrotron radiation experiments stems 
out of the CMS pixel R\&D at PSI, Villigen and it is widely employed on beam-lines world-wide. It offers single photon counting, fast readout, superior PSF 
compared to CCDs, fluorescent suppression by in-pixel energy threshold and no readout noise. The detector is commercially available\footnote{Dectris Ltd., 
Baden, Switzerland.}. A similar development took place also at CPPM, Marseille\footnote{imXPAD SA, La Ciotat, France.}.
Most of the ongoing developments for instrumentation at FELs are also based on hybrid pixel detectors, using silicon as the sensor layer, such as the 
Cornell-SLAC Pixel Array Detector (CSPAD) architecture~\cite{Philipp2011}. Fully depleted CCDs on high resistivity Si~\cite{Holland2003aa} and 
pnCCDs~\cite{Struder2010} are used at light sources and retain some advantages, when frame rate and radiation hardness are not a limitation, as at LCLS. 
Looking towards emerging applications, 3D sensors with active edge appear particularly well suited to the 
fabrication of  micro-dosimeters with a well defined volume at cellular level, a known tissue equivalence factor, and the ability to use in-vivo. 
They are applicable to measurements and control of radiotherapy treatments, such as brachytherapy and hadron therapy. Furthermore, in cancer hadron-therapy, 
the combination of TOF-PET and leading neutrons and gamma detection, possible with 3D sensors, would help the precise determination of the total dose 
delivered to the patient. For these applications Boron- and Lithium-enriched atoms or high Z converters can be used to fill the central region of 3D 
electrodes, thus increasing the total neutron detection efficiency.  

\vspace*{-0.35cm}
\subsection{Monolithic pixels}
\label{sec:3-2}
\vspace*{-0.35cm}

Hybrid pixel detector technology has reached a certain level of maturity and allows unambiguous hit reconstruction in two dimensions.
Bump bonding limits the pitch except for advanced (3D) interconnect technologies enabling chip-to-chip assembly at extremely small pixel 
pitches ($<$10~$\mu$m).

The emergence of monolithic pixel sensors from the late nineties was driven by the necessity to reach material budget and granularity 
performances well beyond the current LHC standards, combined with the capability to cope with high hit rates.
\underline{CMOS active pixel} sensors exploit the industrial CMOS technology for micro-circuits. These also profit of significant industrial 
technology progress for camera sensors, such as OPTO processes. Some of the CMOS fabrication processes include 
the deposition of a lightly doped silicon layer on the standard, highly doped, substrate, prior to the micro-circuit lithography, in order to 
improve the insulation of the micro-circuits from the substrate. The low doping of this "epitaxial layer" favours charged particle detection 
as the electrons generated by traversing ionising particles exhibit a mean free path large enough ($\sim$ 100~$\mu$m) to reach regularly implanted 
sensing nodes with marginal recombination. If the thickness of the epitaxial layer exceeds typically 10~$\mu$m, the signal generated at a 
rate of about 70 $e$-h pairs~$\mu$m$^{-1}$ becomes sufficient to result in a S/N ratio ensuring high detection efficiency.  

Monolithic detectors are considered for very small pixel pitch in the inner layers and also for the cost benefit in outer layers, but for that 
the power per pixel has to be much better than for hybrid detectors. Monolithic pixels offer less mass than hybrid pixel detectors, 
lower $C$, and a single chip solution without bump bonding~\cite{Fossum1997,Turchetta2001}.    The fact that monolithic CMOS pixels combine 
particle detection and the FEE in the same device is a prominent advantage, but turns into a limitation as their industrial manufacturing relies 
on procedures optimised for commercial items which may depart substantially from those needed for charged particle detection. CMOS industry has 
evolved in a direction which allows CMOS pixels to progressively approach their real potential.

Monolithic CMOS pixel sensors in future vertex detector systems rely on the advances in the CMOS technology to cope with the demanding requirements 
of future collider experiments. To this purpose, many R\&D paths are being pursued in the vertex detector community. The shrinking of the feature 
size of CMOS transistors is the most apparent outcome of the evolution of microelectronics. This is being exploited in the design of new pixel systems 
where advanced functions can be performed in the circuitry implemented in each pixel to provide the required signal-to-noise ratio and to handle the 
high data rate. These functions include amplification, filtering, calibration, discriminator threshold adjustment, zero suppression (also called data 
sparsification) and time stamping. 
CMOS sensors for particle tracking were initially designed following the classical guidelines of the NMOS-only 3-transistor cell of image sensors to 
collect the charge generated in an almost field-free epitaxial layer. Thanks to CMOS scaling, more functions are being included in the pixel cell, 
including correlated double sampling and hit binary information from a discriminator. This leads itself to achieve zero suppression at 
the pixel level, which is a highly desirable feature given the large number of channels foreseen and the fast readout of the pixel matrix. New 
strategies are also being devised in terms of the needed readout architecture, introducing a higher degree of parallelism to speed up the transfer of 
information from pixels to the chip periphery. Another way to comply with a high hit rate is to adopt the solutions developed for hybrid pixel sensors. 
This requires that the classical continuous-time analog signal processing chain (charge-sensitive preamplifier, shaping filter) is integrated in the 
pixel cell, together with the digital blocks needed for a high-speed sparsified readout of the matrix~\cite{Traversi:2007}. The implementation of 
complex circuitry in the pixel cell requires a technique to insulate the transistors from the collection volume, to avoid parasitic charge collection. 
This can be done using large deep N-well electrodes as sensing elements or by profiting of the thin oxide insulating layer of the silicon-on-insulator 
(SOI) technology~\cite{Kucewicz2005,Arai2010} to use full CMOS capabilities, without degradation of the charge collection efficiency. 
These technologies make it possible to implement advanced readout architectures, that can be tailored to different applications. In the last few years, 
successful results have been reported  for the 130~nm CMOS technology node~\cite{Neri:2010} and the 200~nm SOI process~\cite{Arai2010}. Recently, the 
180~nm CMOS process featuring a nearly 
20~$\mu m$ thick epitaxial layer with resistivity exceeding 1~k$\Omega \cdot$cm, 6-7 metalisation layers and additional P-wells offers the possibility 
to use both types of transistors inside the pixels thus allowing for a substantial increase of the in-pixel micro-circuit complexity~\cite{Zucca2012}. 
The 90~nm and 65~nm CMOS technology nodes are currently being evaluated as a promising solution to the integration of high density circuitry and high speed 
functionalities in a small pixel. High-voltage CMOS pixels~\cite{peric2011} were pioneered several years ago with a larger feature size 
process~\cite{Kenney1994} and their development is currently pursued on smaller feature size in the perspective of tracker upgrades at LHC.

The use of a moderate- to \underline{high-resistivity substrate} enhances the amount of collected charge in monolithic pixel sensors without compromising 
their competitive thickness. Operating the detector fully depleted has advantages both in terms of signal charge and of collection time, thus improving 
the performance after irradiation by more than an order of magnitude. This can be achieved either with specialised CMOS processes featuring a resistivity 
in excess of 1~k$\Omega \cdot$cm, or with the SOI process on high resistivity handle wafer. 
Both approaches have been successfully exploited in monolithic pixel R\&D ensuring S/N values of 30-40 from minimum ionising particles at room temperature. 
Sensors based on high resistivity CMOS pixels thinned to 50~$\mu$m, are installed in the beam telescope developed within the EUDET EU project~\cite{Aguilar:2012ez}. 
They provide a $\sim$ 3~$\mu m$ single point resolution at a rate of $\sim$ 10$^4$~frames~s$^{-1}$ over an active surface of about 2~cm$^2$ paved with 670,000 
pixels and can cope with particle rates exceeding 10$^6$~cm$^{-2}$~s$^{-1}$.
Based on a more specialised technology, \underline{DEPFET pixels} have been developed since the late 80's~\cite{Lutz1987} and integrate a p-MOS transistor in 
each pixel on the fully depleted bulk. Electrons, collected in the internal gate, modulate the transistor current. Their low input capacitance ensures low noise 
operation and makes them suitable for a broad range of applications from collider detectors~\cite{Rummel:2010} to X-ray astronomy~\cite{Stefanescu2010aa}.

It is worth mentioning that, besides device scaling, advanced CMOS technologies bring along other desirable features that pixel R\&D projects are presently 
exploring. Metal interconnections are isolated by low-$k$ dielectrics to reduce parasitic capacitance and time constants of transmission lines. This may be 
exploited by minimising the electronic functions in the pixel cell and by connecting each pixel to signal processing blocks in the chip periphery. Thanks to 
the very high integration density of the CMOS process, this peripheral readout strategy may require just a small silicon area outside the pixel matrix itself. 

Monolithic pixel sensors are being successfully applied to several fields of imaging. In \underline{electron microscopy}, cameras based on a monolithic CMOS 
pixel sensors are becoming commercially available. In particular, a reticle size 16~M pixel sensor with 5~$\mu$m pitch and a frame rate up of 400~Hz, 
originating from the linear collider vertex R\&D, has been jointly developed by LBNL, UCSF and Gatan Inc.\ and it is currently available as a 
commercial product\footnote{Gatan Inc., Pleasanton, CA, USA.}.  For \underline{soft X-rays}, where single photon energies are below hybrid pixel thresholds, 
low noise, monolithic detectors are being developed. Here, challenges include low noise, and thin entrance windows in order to ensure high quantum efficiency 
at low energies. Aiming to scientific applications at XFELs, the PERCIVAL (Pixellated Energy Resolving CMOS Imager, Versatile and Large) 
project developed by a collaboration of RAL, DESY and Elettra uses monolithic CMOS technology to achieve a frame rate in excess to 100~Hz for full frame 
readout, dynamic range from 
1 to 10$^5$ photons, single-photon counting capability with low probability of false positives in a mega-pixel detector.
CMOS sensors are becoming attractive in intra-oral dental radiography where they offer several advantages including dose reduction and are commercially 
available\footnote{Schick Technologies Inc., Long Island City, NY, USA.}.

The low material budget, high granularity and the capability to deal with large particle fluxes afforded by CMOS pixels makes them well suited as high 
performance devices in \underline{beam monitoring}. Sensors developed for the EUDET telescope and the STAR HFT detector are being prepared as a new generation 
of hadron beam monitors replacing gas chambers and in telescopes for fast on-line proton imagers.

\vspace*{-0.35cm}
\subsection{Interconnection}
\label{sec:3-3}
\vspace*{-0.35cm}

The interconnection between the sensor and the front-end ASIC in present hybrid silicon pixel detectors is provided by micro-bump bonds with typical 
diameters of 25~$\mu$m and pitches of as low as 50~$\mu$m. Each bump bond connects one pixel cell on the sensor to the corresponding pixel cell in the 
front-end ASIC. The current \underline{bump bonding} technique is based on full wafer processing, which requires the use of handle and support wafers in case of thin 
wafers. It involves a series of processing steps: deposition of under bump metalisation layers, electroplating or evaporation of the actual bump material, 
photo-lithographic and etching steps, tacking and possible re-flowing of the bumps. Bump bonds used for hybrid pixel detectors installed in high energy and 
nuclear physics experiments or in medical imaging are based on Pb-Sn, Ag-Sn or indium bumps. ASICs are thinned to 150~$\mu$m to minimise the material in 
tracking detectors, though prototype developments for future detectors already target ASIC thicknesses of less than 100~$\mu$m. The thicknesses of 
present detectors are as low as 150-200~$\mu$m. 

For the \underline{LHC upgrades}, bump bonding is still the interconnection technology of choice, since the pitch is not yet critical and the bonding can be 
performed by industry. For pixel pitches below 50~$\mu$m, industrial processes are available but not readily accessible for the research 
community and bump bonding is still one of the main cost factors for hybrid pixel detectors. Bump bonding is usually provided by small- to medium-size 
private and semi-private companies which are ready to process small wafer quantities and are available to perform process development together with research 
customers. The development of cost-effective and readily accessible bump-bonding facilities, suitable for processing small numbers of prototype sensors 
would be a great advantage for the community. Such an initiative has been proposed, for example, in 
the UK national laboratories (CCLRC RAL/Daresbury), and this or similar projects, would have the support of the experimental collaborations.
Extending the use of hybrid pixel detectors to larger radii will require lowering of the cost of fine-pitch bump bonding. Industrial scale solutions at 
potentially lower cost are already available for larger pitches, from about 100-200~$\mu$m or larger. A further cost reduction can be achieved by using 
chip-to-wafer or wafer-to-wafer bonding by thus reducing the number of processing steps and manual handling.
Novel techniques, which will also allow reducing the pitch to less than 20~$\mu$m, are under study for future pixel detectors. Direct metallic connections, 
e.g.\ using Cu pillar bumps and SLID structures~\cite{Enquist2009,Macchiolo2011}, will allow us to achieve fine pitch connections, while requiring a very high 
planarity of the components and limited possibility of rework. 

Technologies such as \underline{through silicon vias} (TSV) are being studied to fan out the contacts for a chip to the back side of the ASIC instead of 
routing them to the wire bonding pads located on the edges of the chip~\cite{Macchiolo2011,Gonella2011aa}. 
TSVs have been successfully applied to FE-I3 chips which have been operated from the backside~\cite{barberoTSV}. 
A redistribution layer on the ASIC back side can be used to bring the electrical contacts to a matrix sufficiently large to use standard Ball Grid Array (BGA) 
type connections. This will allow forming very compact modules by omitting the space required for the wire bonding 
connections. This will allow to form four-side buttable assemblies thus reducing the insensitive area between adjacent chips. The need of future pixel 
detectors to further reduce the material budget and thus use thinner wafers is fully in line with the requirements for making TSVs by considering that the 
limitations are given by the aspect ratios of the technologies used to generate the holes.
BGA type connections can also be used for monolithic silicon pixel detectors without the requirement to form a TSV. The pads for the connections can be 
integrated on the front-side of the chip. However, in conjunction with using very thin silicon components, which are connected via standard BGA soldering 
techniques the effects of thermal stress, will be more pronounced.

\underline{3D integration} technologies promise to provide a very elegant way of implementing advanced readout architectures of a pixel matrix~\cite{Yarema2010}. 
Digital electronics can be removed from the silicon layer where the sensing electrodes and the analog front-end circuits are located, eliminating 
interference due to substrate coupling. In a purely digital layer, the designer is allowed to integrate various advanced readout concepts, such as 
time-ordered readout of hit pixels, in a ``data push'' or triggered way, enabling a pixel matrix to deal with a hit rate of 100~MHz~cm$^{-2}$ with 
a readout efficiency close to 100\%~\cite{Gabrielli2011}. 
There are several approaches to 3D integration, which differ in terms of the minimum allowed pitch of bonding pads between different layers and of 
vertical TSVs across the silicon substrate. The HEP community focused a large effort on the design of 3D integrated circuits with the ``via first'' 
technology, where the TSVs are etched in the silicon wafers in the early stages of CMOS fabrication, provided by Tezzaron/GlobalFoundries~\cite{Yarema2010}. 
Even though this has only led to a fully functional device after several cycles, it remains a very promising approach. However, even with less aggressive 
3D technologies, 
such as the so-called ``via last'' processes where TSVs are fabricated on fully processed CMOS wafers, a significant advantage can be gained in most 
cases, as only one or two connections are needed between the analog and digital blocks of a single pixel cell, and the digital layer can use 
low-density peripheral TSVs to reach backside bonding pads for external connection. Devices based on the ``via last'' technique are being studied 
within the AIDA EU project.

Sensor stitching opens up a promising path to progress towards thinner modules, with clock and signals routed on metal lines in the chips and larger area 
sensor arrays. Stitching is an industrial standard offered for several processes of interest for monolithic CMOS pixels. Experience with sensor stitching 
is currently limited to relatively large feature size processes~\cite{Zin2010} and more work is needed. 

The development of interconnection technologies will have to address several key issues for new generations of pixel detectors. New technology and process 
options for the typical volumes used in experiments might reduce the cost for bump bonding of pixel detectors and thus would allow to construct larger area 
detectors. At the same time a further reduction of the overall material is the focus of most projects. The complexity in the handling and processing of thin 
wafers increases. The access and exploitation of 3D technologies will enable building even more compact pixel modules and will go hand in hand with new 
packaging technologies. However, research applications will be dependent on industrial developments for access to the via-first or via-middle processes 
involving steps which must take place within the CMOS foundries.

\vspace*{-0.35cm}
\subsection{Alternative semiconductor materials}
\label{sec:3-4}
\vspace*{-0.35cm}

The introduction of new materials for pixel sensors is motivated by two main reasons: increased radiation tolerance for HEP and 
increased attenuation coefficient for photon imaging. 

\underline{Diamond} has excellent radiation tolerance and its larger band-gap compared to Si reduces
the leakage current. This comes to the price of a $\sim$0.5 smaller number of charge carriers generated in the same 
equivalent thickness expressed in radiation length units. Diamond detectors have been extensively used in beam monitors 
exposed to the most severe background conditions at colliders. The steady luminosity increase at colliders, in particular 
at the LHC, justify their consideration as a replacement of Si in the innermost layer(s) of the vertex 
trackers~\cite{Asner2011aa}. The RD42 collaboration promotes the development of diamond pixels for vertex tracking 
applications. A single crystal CVD diamond pixel sensor bump-bonded to an ATLAS pixel chip has demonstrated efficiency 
in excess to 99.9\% and single point resolution of (8.9$\pm$0.1)~$\mu$m on a 50~$\mu$m pitch, in a beam 
test~\cite{Velthuis2008aa}. Diamond detectors are being considered by CMS and LHCb for their vertex tracker upgrades. 

\underline{SiC} has been considered for its attractive properties which include low leakage, seven times higher breakdown field 
compared to Si, two times higher carrier saturation velocity and radiation tolerance. In addition, it is blind 
to visible light, which is advantageous for several applications. Thanks to the recent progress in wafer manufacturing, 
very-low noise SiC pixel sensors can be manufactured~\cite{Caccia2011aa}. The results are interesting in terms 
of the achievable energy resolution and the broad range of operating temperature. 

At energies above about 15-20~keV, Si sensors, of the few hundred micron thickness generally used, begin to become nearly 
transparent. For harder X-rays, developing hybrid sensors with higher Z materials will be important.
Hybrid pixels matching \underline{high-Z materials} (GaAs, CdTe, CdZnTe) as sensitive layer with readout chips derived from HEP technology 
are well established in imaging applications based on single photon counting, such as medical radiology and crystallography 
at light sources. This is especially important for FEL applications since these are planning to use higher harmonics, up 35~keV, 
and Si would result in poor quantum efficiency as well as increased radiation doses in the underlying ASICs. 
While these materials may not be necessarily relevant for HEP applications, they make possible the porting of readout chips derived 
from the HEP R\&D to spectroscopic applications requiring high energy resolution and photon energy sensitivity beyond 10~keV. 

\vspace*{-0.35cm}
\section{Electronics, Services and System issues}
\label{sec:4}
\vspace*{-0.25cm}

\vspace*{-0.25cm}
\subsection{Read Out Electronics}
\label{sec:4-0}
\vspace*{-0.35cm}

Pixellated high-resolution semiconductor detectors and the revolution they have enabled in particle tracking have been made possible primarily through 
the parallel development of new detectors and read-out electronics through high-density microelectronics technologies. 
The present trend in the pixels deployed in the LHC experiments is to include a significant amount of electronics with transistor counts 
well beyond 1000 per pixel, and a power consumption of several 100~mW per cm$^2$ of active area. The presently installed LHC readout chips 
are manufactured in 250~nm CMOS technologies using special layout and design techniques to increase the radiation tolerance both in terms 
of total dose and single event effects~\cite{Anelli1999aa}. 
More advanced CMOS technologies offer reduced feature size and increased robustness to total dose often eliminating the need for enclosed 
layout transistors. Pixel sizes are therefore likely to be less limited by circuit density than by interconnection technology. 
\underline{Power consumption}, \underline{material budget} and \underline{cost} will therefore be more significant constraints as technology 
for smaller pixel pitches is likely to be available.

For the SuperB project the main requirement for the FEE is to cope with the high hit rate of 100~MHz~cm$^{-2}$ with in-pixel sparsification and 
fast time stamping to better than 1~$\mu$s in order to keep the bandwidth per module below 5~Gbit~s$^{-1}$.
The upgrade of the ALICE inner tracker system, the STAR HFT at RHIC and the vertex trackers at lepton colliders have to limit the material 
to new levels and therefore consider thinner detectors, with less signal and hence more analog power consumption, for a different reason. This has 
to be considered carefully as a low mass system requires the power and signal distribution, cooling and mechanical structure to be light and 
therefore require low power consumption~\cite{Spieler2010}. Several projects envisage powerful integrated cooling solutions such as micro channel 
cooling (see \ref{sec:4-1}) and can tolerate power consumptions above several hundreds mW~cm$^{-2}$. Lepton colliders with more extreme material budget 
requirements ($\le$0.2$\% X_0$ per layer) foresee air cooling with a power budget of $\sim$100~mW~cm$^{-2}$ or less.  Power pulsing 
is proposed given the favourable beam structure at a linear collider, with specific electronics blocks of the front-end ASIC being switched off 
in the time between bunch trains. This is not trivial at component nor system level and will require extensive study. Power pulsing cannot be 
applied at the LHC due to the continuous bunch structure. To further reduce material electro-mechanical integration and interconnect are under 
extensive study. One area of research are silicon through vias (TSVs) which allow a direct connection from the bump bonded read-out ASIC to the 
back side of the silicon. This avoids wire bonds, increases geometrical acceptance and allows a leaner electro-mechanical architecture.

In addition to performance parameters like data rate, time resolution and required on-chip data processing, the pixel size and material 
budget are also ultimately constrained mostly by power consumption. The analog power consumption for a given readout speed and S/N ratio is 
strongly dependent on the $Q/C$ ratio, i.e.\ the collected signal charge $Q$ divided by the capacitance $C$ of the collection electrode. 
This ratio represents the signal strength at the input, and has to be compared to the noise of the input transistor to obtain the S/N 
value. Typically the thermal noise is dominant in HEP pixel systems. Since the thermal noise is only weakly dependent on the current 
(square root in weak inversion to 4$^{th}$ power root in strong inversion), this and hence the power needed for a certain S/N will vary strongly 
with the $Q/C$ ratio and be inverse proportional to $(Q/C)^n$, with 2$\le$n$\le$4. It is therefore important to compare different technologies 
in terms of their $Q/C$ ratio as this is key to lower analog power.

For \underline{monolithic pixel} sensors collecting charge from just 10-15~$\mu$m thick epitaxial layer, the $Q/C$ ratio and power consumption might not 
be better compared to hybrid pixel detectors. The reduction in power consumption and $Q/C$ improvement can be achieved by minimising $C$ 
decreasing the size of the collection electrode and by increasing the charge collection and depletion depth by applying reverse bias voltage 
with a moderate- to high-resistivity substrate. Operating them in depleted mode will also significantly increase radiation tolerance as charge 
is collected mainly by drift and not by diffusion.
Analog and digital power consumption typically differ by less than an order of magnitude in pixel detectors in high energy physics. Analog power 
reduction by an improved $Q/C$ or by power pulsing has therefore to be matched by a similar reduction in digital power. Power pulsing schemes 
also for the digital read-out are considered. However also architectures limiting the off-chip data rate and minimising the number of 
operations are under study. As an example, in the LePix project~\cite{Mattiazzo2012}, the in-pixel circuitry is reduced to minimise $C$, with data 
from different pixels combined in different projections and immediately transmitted to the 
periphery. The different projections are chosen carefully to minimise ambiguities and for a practical example with 4 projections, simulations 
indicate that occupancies of ~ 50~hits/cm$^2$ can be handled while only 4$N$ signals need to be treated at the periphery for an 
N$\times$N pixel matrix. 
An important component of the power consumption is the off-detector data transmission, and any progress to optimise power/data rate in this 
area will be beneficial to almost all new applications. Dedicated architectures are being developed, addressing directly the conflict between 
high pixel granularity and fast read-out requirements. They are based on multi-layer devices, where one is dedicated to
charge collection while the others are devoted to signal processing.

The choice of a simple versus a more complex pixel architecture appears also in the design of pixels with a \underline{time} \underline{resolution} 
beyond that required 
by the LHC 25~ns bunch crossing and its upgrades.  The LHCb upgrade is contemplating proton taggers situated a few 100~m up and downstream of the 
experiment with $\cal{O}$(1~ns) time tagging within the pixel.
Examples featuring a time resolution in the few ns range are 
TIMEpix~\cite{Llopart2007aa,Poikela2012aa}, ToPix~\cite{Mazza2012aa}, Chronopix~\cite{Baltay} and the TDCpix~\cite{Noy2011aa}. Distributing a 
time counter value over the pixel matrix and attaching its value to the hit information is an excellent way to produce time information at the 
few ns level, but the time reference distribution introduces higher power consumption and additional digital noise in the analog front-end. 
This was avoided in an alternative approach by replacing the digital time bus with a local oscillator~\cite{Poikela2012aa}. For the lepton 
colliders a time binning of several nanosecond range might be required to reduce the occupancy due to beam-induced backgrounds. For 
sub-nanosecond time resolution, varying signal amplitude produced by the sensor and resulting discriminator time walk need to be compensated. 
An approach recently employed uses a constant fraction discriminator (CFD) and TDC in each pixel of the matrix with data sent out on a digital 
bus~\cite{Dellacasa2011aa}. 
Another approach employed pixels with a single threshold discriminator sending the discriminated signal to the periphery of the read-out 
ASIC to register the arrival time and time-over-threshold information to compensate for time walk~\cite{Noy2011aa}. 
The first approach has more complex pixels with higher analog power consumption due to CFD and TDC. The second consumes more in the digital part 
at the chip periphery, offering lower consumption in the pixel matrix and avoiding clock distribution over the sensitive pixel matrix. 
Time resolution of less than 200~ps has been demonstrated with a particle beam~\cite{Noy2011aa}, and recently shown to be limited by the sensor. 
Investigations on the sensor are needed to assess whether this time resolution can be improved.

New challenges provided by future applications are due to the addition of \underline{trigger capabilities}. Significant physics improvements can be achieved by 
building track elements and reconstructing vertices from the pixel tracker information at level-2, or even level-1. Difficulties are related to the complexity 
of the interconnections between different layers, required by the trigger functionality, together with power and material budget issues. 
This is being investigated by several experiments. In particular, in the LHCb upgrade triggering would be performed in a fully-flexible software trigger, 
which can thus utilise displaced vertex triggering at the first level. This requires a high \underline{output rate} from the FEE, and the output rate is a 
limiting factor in the instantaneous luminosity and proximity of the sensors to the beam. The output rate of the chip is currently limited by the number 
and speed of output stages which can be installed, together with their power consumption, which is currently 50~\% of the pixel chip budget.  However, 
greater data output speeds would provide flexibility to go to higher luminosities and closer to the beam lines and would be a great advantage.  The hottest 
pixel chips closest to the beam will see rates of approximately 500~MHz pixel hits and will output $\simeq$~12~Gbit~s$^{-1}$. The total data rate is 
approximately 2.8~Tbits~s$^{-1}$. Achieving this output rate requires the implementation of on-chip zero suppression and the construction of 4$\times$4 
super-pixels in the pixel chip. 

The recent development of \underline{FELs} facilities, producing ultra-intense, ultra-short pulses with a peak brilliance increase over conventional third 
generation synchrotron radiation machines by nine orders of magnitude make at the same time  possible and necessary  to perform single shot imaging experiments. 
These prevent photon counting architectures, and require detectors integrating the photons received in a single pulse. Still they must keep the total noise 
well below a single photon and operate with a very large dynamic range, up to 10$^6$. Currently operating hard X-ray FELs operate at $\sim$100~Hz frame rates. 
Future FELs, based on superconducting accelerators, will operate at orders of magnitude higher pulse repetition rate, requiring much faster readout together 
with more sophisticated data handling. Contrary to particle physics experiments where the occupancy is low and sparsification is possible, these applications 
typically require full frame readout. Various architectures are presently being developed, including dynamic gain 
switching (AGIPD)~\cite{Henrich2011,Potdevin2011}, multiple parallel gains (LPD), or non-linear gains (DSSC)~\cite{Porro2012}. 
Full frame applications, such as tomography, dynamic imaging and scattering experiments are able to take advantage of higher frame rates as 
brightness increases. Photon correlation spectroscopy will have increasing data rates and will require a more precise time resolution.

\vspace*{-0.35cm}
\subsection{Cooling}
\label{sec:4-1}
\vspace*{-0.35cm}

The cooling system required to remove the heat dissipated by the readout electronics and heavily irradiated Si sensors represents a major contribution 
to the overall material budget and defines an important constraint on the FEE design and functionalities.
In order to minimise the material budget new approaches to cooling are being pursued. The phase-1 CMS pixel upgrade adopts CO$_2$ cooling based on ultra 
light mechanics and thermal pyrolytic graphite (TPG), a  very light material with excellent in-plane thermal conductivity ($>$1400~W/km), for the 
substrate. The contribution of coolant and pipes to the material budget will decrease from 2~\%~$X_0$ of the current detector to 0.1~\%~$X_0$.

Designs aiming to ultra-light ladders have so far relied on \underline{passive cooling}. The HFT project at STAR has pioneered the study of thin 
CMOS monolithic pixel ladders where heat is extracted by air-flow. Tests showed that a flow speed of 10~m~s$^{-1}$ removes up to 0.2~W~cm~$^{-2}$ with an 
acceptable temperature rise above ambient and displacements due to vibrations below 6~$\mu$m r.m.s.\ on a 200~mm-long ladder~\cite{Greiner:2011aa}. 
Air cooling is also adopted by BELLE-II using $\le$1~m~s$^{-1}$ flow.
The power dissipation constraint for passive airflow cooling has major implications on the FEE design. In order to overcome 
this limitation more innovative approaches are needed such as the integration of the cooling system into the CF support structure or the Si wafer itself 
using \underline{micro-channel} evaporator. The first approach is developed for the SuperB project, to absorb the 1.5~W~cm$^{-2}$ dissipated by the pixel 
front-end electronics and by ALICE for 0.5~W~cm$^{-2}$. In these design, the support with integrated cooling is build with carbon fiber micro-tubes, with 
an hydraulic diameter of about 200~$\mu$m obtained by poltrusion process and polyimide tubes with CF filaments, respectively. Measurements on the SuperB 
support prototypes, with a total material budget as low as 0.11\% X$_0$, indicate that 
it can remove {\cal{O}}(1~W~cm$^{-2}$) over a length of tens of cm~\cite{Bosi:2011}. The integration of micro-channels in the Si wafer itself was originally 
developed for cooling microprocessor chips, where it can remove {\cal{O}}(100~W~cm$^{-2}$) on small surfaces~\cite{Thome}. It allows the integration of the 
cooling system within the detector structure itself, with obvious advantages on the optimisation of thermal bridges and transparency to incident 
particles~\cite{Mapelli2012}. Micro-channel cooling is currently being evaluated also by the LHC collaborations and NA62~\cite{Mapelli2011}

\vspace*{-0.35cm}
\subsection{Power Distribution and Data Transmission}
\label{sec:4-2}
\vspace*{-0.35cm}

Power distribution and data transmission are becoming important system aspects in the design and optimisation of vertex trackers.
Due to the low voltage required by deep submicron microelectronics the voltage drop along the low-mass cables is significant. Power 
distribution for the next generation of detectors is thus considering using low currents and high voltages through serial powering~\cite{Stockmanns2003} 
or DC-DC converters~\cite{Feld2011}. The former appears advantageous in terms of material budget~\cite{Gonella2011aa}.

The transmission of data from the on-detector front-end electronics to the data processing off-detector is becoming a major challenge. The bottleneck in bandwidth 
is in the transit region from around 10~cm up to a meter.  In the front-end electronics the high bandwidth is achieved with parallel transmission of data and 
transmission and over long distances is made with powerful optical serial links. Today the large amount of services required to power, cool and read out modern 
semiconductor trackers compromises their performance. 
\underline{Wireless data transmission} is today the dominating technology for local data communication. A growing 
interest in 60~GHz technology worldwide is driven by the large unlicensed bandwidth offered by the technology~\cite{Yong2007}. An increasing number of 
components are being developed and are already commercially available. Due to the short wavelength, their dimensions are compatible with the 
sizes of detector components.  The 60~GHz electronics can be processed with the same CMOS technology used for front-end electronics for pixel trackers~\cite{Marcu2009}.
One of the main advantages is that the data path is not fixed to the routing of wires. This opens for topological readout of the vertex tracker~\cite{Brenner2010}. 
However, the bandwidth of a single wireless network at 60~GHz is far too low and a large number of wireless links is required. 
The operation of densely packed wireless links with minimal interference imposes the main challenge to the usage of the technology in future detectors. 
Developments are required in areas of very directional antennas, data modulation technology and in confinement of power and networks. To make full profit of the 
technology, methods to cross boundaries not transparent to millimetre waves need to be developed.   
   
\vspace*{-0.35cm}
\section{R\&D cycle from HEP to spin-offs back to HEP}
\label{sec:4-3}
\vspace*{-0.35cm}

The interplay between HEP and imaging applications can be understood as a two-way process by following the development of one of the first project 
to successfully transferring architectures originally developed for particle physics detectors to medical imaging and now back to LHC upgrade 
applications. The Medipix collaborations, which consist of a large number of institutes, are dedicated to the development of general purpose 
pixel detector readout chips mainly used for noise free particle detection and counting. This activity originated from the LHC-1 readout chip~\cite{lhc1}, 
which was successfully applied in the WA97~\cite{wa97} and NA57~\cite{na57} experiments and became the precursor of the readout chips for the Delphi 
VFT~\cite{vft} and then the ATLAS and ALICE pixel FEE chips. The Medipix2 focused on developing chips at the 250~nm CMOS technology node, used by the 
current LHC pixel readout chips,~\cite{Campbell2011} and the Medipix3 develops chips in 130nm CMOS. The Medipix2 chip~\cite{Llopart2003} was 
initially foreseen for applications in X-ray imaging and, indeed, it has been used successfully for many experiments in the field of X-ray radiography 
successfully transferred to industry\footnote{PANalytical B.V., Almelo, The Netherlands}, 
marketing the chip for commercial X-ray diffraction cameras. Other applications include science 
at FEL facilities~\cite{Pennicard2011}, neutron 
detection and radiography, electron microscopy, wavefront sensing in adaptive optics for astronomy, and radiation monitoring and dosimetry. 
It has also been used extensively to study and characterise new semiconductor sensor materials and structures. Particle and nuclear physics 
applications of the chips include nTOF, ISOLDE, UA9 and, more recently in the ATLAS-MPX system~\cite{Vykydal2009} providing independent, 
real-time information about the radiation environment in the experimental cavern. The Timepix chip~\cite{Llopart2007aa} was developed from the 
Medipix2 design at the request of and with funding from the EUdet consortium. The readout electronics was modified to measure the particle arrival 
time with respect to an externally applied shutter signal and Time-over-Threshold (ToT) information. These modifications made it an extremely 
versatile solution for the development of new detector technologies and for beam hodoscopes such as that assembled by LHCb~\cite{Akiba2012}.

The Medipix3 is a new chip~\cite{Ballabriga2011}, which uses inter-pixel event-by-event charge summing and hit allocation to extract energy sensitive 
images from an otherwise dispersive semiconductor detector medium. 
The collaboration has been so far successful in attracting new members and in technology transfer. 
The Timepix3 chip has a data driven readout architecture, which makes it an ideal prototype for the LHCb VELOpix upgrade. Its architecture is 
also being evaluated for the Alice SPD upgrade. It can be expected that slightly modified versions of this chip could be developed for either 
experiments. Further developments will explore smaller pixel pitches and a larger pixel matrix, incorporate a new architecture to support power 
pulsing to provide a valuable test vehicle for linear collider detector engineering studies. Through the pooling of resources, it has 
been possible to make significant strides forward in the development of pixel readout electronics during the time when the LHC detectors were under 
construction and commissioning. The involvement of European industries has made possible to maintain the support of the community requests. 
Key sensor and interconnect technologies needed by the HEP community could be sustained during the 'trough' between the procurement phases of the 
LHC experiments. In this respect, it is interesting to consider a 'generic' pixel detector for an LHC experiment and compare its hit rate and readout 
requirements with other applications. In a pixel barrel of 1~m length with a radius of 50~mm and 10$^4$ hits per 25~ns BCO, the 
rate of hits per mm$^2$ is $\sim$1~MHz. This number is comparable to the track hit rate foreseen on the innermost chips of the LHCb 
VELOPix. However, given the geometry and size of the barrel it is unlikely that all hits will be read out without some form of data selection.  
For comparison, in a medical CT system the hit rate of the direct beam (which does not cross the body of the patient) is about $\times$1000 higher. 
Of course, there is no readout system available at present for such a data rate but one can see that the requirements of HEP and medical imaging 
are at least comparable. The evolution of the Medipix programs clearly demonstrates how the collaboration with scientists from outside of HEP can be 
mutually beneficial.  Access to deeper sub-micron CMOS will involve engineering runs at increasing costs making it very difficult, if not impossible, 
for individual experiments to have 'bespoke' chip designs. Collaboration and compromise on specifications may become mandatory and the experience of 
the MediPix collaborations may provide an important template for such future efforts. 

\vspace*{-0.35cm}
\section{Test Facilities}
\label{sec:5}
\vspace*{-0.35cm}

The large number of R\&D pixel projects technologies under development for applications from imaging detectors for photon science applications to 
highly granular thin pixels for future linear colliders, require adequate test facilities. Special needs are set for high 
occupancy studies for HL-LHC pixel detectors and special beam structures for LC pixel detectors.
The characterisation of pixel sensors typically entail the study of a number of properties, including charge-collection efficiency, spatial resolution and 
detection efficiency before and after irradiation, which require a test beam facility. High momentum particles are preferred since they minimise the effects of 
multiple scattering. Most R\&D groups carry out regular beam tests at FNAL, DESY or CERN accelerators since high energies beams 
are of major interest. In recent years, the availability of beam time has been limited but sufficient. In 2013 this situation will 
change with the CERN accelerator long shut down for the upgrade work. 
Only facilities at DESY, FNAL and SLAC will be available. FNAL will commission a second test beam line to be able to accommodate a larger number of users. 
In the years after the SPS long shutdown the number of requested weeks at test beam facilities will not be reduced compared to the current numbers. The extensive 
test beam program for HL-LHC detectors to test new sensors and later full prototypes, will continue with an increasing demand and urgency. Also LC-related tests will 
continue their multiple efforts, in the next several years, implying that an increase of requested beam time is probable as ILC detectors will focus on more system 
tests with combined detectors and CLIC-specific activity will ramp up. The pixel groups of the b-factory experiments, of PANDA~\cite{Calvo2010} and CBM~\cite{Heuser2007} 
at FAIR will also need test beam time. It is essential that the number of test beam facilities world-wide will not be reduced. 

For HL-LHC pixel detectors, high rate tests, possibly with beam particles, are needed with intensities of the order of 2$\times10^8$ particles cm$^{-2}$.
Currently the only facility providing rates reaching this level is the H4irrad facility at the SPS, which is heavily subscribed. DESY and FNAL are studying the 
possibility to provide high rate test facilities emulating occupancies comparable to the HL-LHC conditions. Since more facilities are needed to accommodate the demands 
of the community, these new high rate facilities are highly desirable.  
Lowering the power consumption is one of the key challenges for pixel developments towards LC detectors. The baseline approach to control the 
power consumption of LC detectors is power pulsing, A LC-like spill structure would be useful to study power pulsing schemes. 
For HL-LHC bunch-crossing frequencies of 20~MHz or 40~MHz are currently under discussion. 
To enable the study of efficiency versus the phase between on-detector and bunch-crossing clocks for future HL-LHC detectors, beams with a 25~ns bunch structure 
are needed to determine the necessary timing precision for the readout electronics and optimise timing through the entire readout chain. 

Technical infrastructures already available at the beam areas such as beam telescope, magnets and standardised CO$_2$ cooling systems reduce the user set up 
time and increase the number of groups per test beam period. With the EU funded efforts (EUDET and AIDA), and the strong support of the hosting labs, the 
overall infrastructure at the test beams at CERN and DESY were significantly improved in recent years. The test beam line at FNAL had useful infrastructure 
implemented. Additionally the collaborations added "private" infrastructure to different beam lines. As a result, good beam telescopes are now available 
(Timepix, AIDA telescope and its copies) but more would be appreciated by the community. Besides high spatial resolution, good time resolution is desirable. 
Some groups have expressed interest for more generalised data acquisition systems for all the different beam telescope (as in the EUDAQ project). HL-LHC 
pixel sensors need to be cooled to temperatures as low as -35~$^o$C during the tests, especially after high irradiation. Significant effort is put into 
providing adequate cooling systems. The adoption of standardised CO$_2$ cooling plants,  similar to those already available for the Timepix telescope and planned 
for the AIDA infrastructure, appear to be helpful. Magnets for the study of detector performance in strong magnetic fields are also required. Different magnets are 
available but typically their field does not exceed 3~Tesla or the bore is not large enough to accommodate the systems under test.

\vspace*{-0.35cm}
\section{Conclusions}
\label{sec:6}
\vspace*{-0.35cm}

Semiconductor detectors for vertex tracking have provided the much needed technological breakthrough to ensure the LHC detectors enjoy the same accuracy in track 
extrapolation as the last generation of $e^+e^-$ collider experiments, in a much harsher and demanding environment. Driven by 
track density and single point resolution requirements, the future of sensors for vertex tracking seems to belong entirely to pixellated sensors with point 
resolution $\le$ 10~$\mu$m, time resolution of the order of few to 25~ns and tolerant to ever increasing radiation levels. The fast pace of development of industrial 
semiconductor processes provides us with opportunities and challenges for the ongoing R\&D towards LHC upgrades and future collider projects. Advanced CMOS 
processes offer reduced feature size and increased robustness making pixel pitch less limited by circuit density than by interconnection technology. 
3D layout, charge multiplication, very slim edges and monolithic technologies are pushing the R\&D into new domains in terms of radiation tolerance, pixel pitch 
and material budget. The turn around times and costs for bump bonding of hybrid pixel sensors at decreasing pixel pitch can be a limiting factor. 
New interconnection techniques promise to provide ways for implementing advanced read-out of the pixel matrix.
However, HEP instrumentation is not in the position of fully exploiting cutting edge commercial technologies. Its R\&D is presently evaluating technologies 
which are several generations behind state of the art microelectronics and the gap is likely to increase, in the absence of adequate R\&D resources.
The length of the development and construction cycle brings the need to match the technology lifetime and identify suitable established technologies. Mask costs 
for modern CMOS processes grow dramatically as feature sizes shrink, 3D multi-tier interconnection requires a long and expensive learning phase, making prototype 
development a more challenging and concentrated activity. New schemes must be developed to enable HEP R\&D, performed typically by small research groups with limited 
budgets, to be aggregated and organised in larger coherent teams. New collaborative structures are mandatory to increase the capital investments, strengthen engineering 
competences and minimise risks with collegial reviews, compensating for the higher investment costs required to access advanced technologies.
The characteristics of the R\&D cycle dominating particle physics with long development times alternating with experiment construction present challenges and require 
experiments and R\&D collaborations support the diversification of applications between particle physics and other domains of science. Several of these applications 
share much of the requirements and challenges of particle physics. Also smaller scale applications such as b-factories and nuclear science experiments boost R\&D towards 
performances essential for use at future larger colliders. Coming into operation in the next few years, they will add much needed especially experience with new 
technologies and system aspects. In order to keep the pace of the evolving requirements new concepts have to be constantly introduced, not only in the development of 
sensors, but also in readout electronics and services, in particular support structures, power distribution, power dissipation and cooling systems.
The technology transfer process between particle physics and other applications works both ways and we have examples of returns from imaging to collider applications. 
It is important for the community that the number of test beam 
facilities is kept not below their current level and that of high rate test facilities is increased to make sure that the R\&D can continue at a vigorous pace.
 
\section*{Acknowledgements}

We are indebted to G.~Bolla, G.~Casse, B.~Di~Girolamo, M.~Mikuz, C.~Rembser, H.~Sadrozinski, M.~Swartz, N.~Unno and M.~Vos for 
their contributions. M.~Caccia, C.~Damerell, M.~Demarteau, E.~Heijne, A.~Marchioro, H.G.~M\"oser, A.~Seiden and N.~Wermes have reviewed the 
document and provided us with many precious inputs and suggestions. 
We are grateful to A.~Cattai and the ICFA Instrumentation Panel for strongly supporting the preparation of this report.

\bibliographystyle{elsart-num}
\bibliography{sidet_v2}

\end{document}